\documentclass[
 reprint,
 preprintnumbers,
 amsmath,
 amssymb,
 aps,
 prstab,
 floatfix,
]{revtex4-1}

\usepackage{graphicx}			
\usepackage{verbatim}			
\usepackage{wasysym}			
\usepackage{mathtools}			
\usepackage[makeroom]{cancel}		
\usepackage{dcolumn}			
\usepackage{hyperref}			
\usepackage{color}			

\newcommand{\pd}{\partial}				
\newcommand{\dd}{\mathrm{d}}				

\newcommand{\Qb}{ Q_{\beta}}		

\newcommand{\omsc}{\omega_\mathrm{sc}}

\newcommand{\Omk}{\Omega_k}

\newcommand{\kbar}{\bar{k}}

\newcommand{\Lth}{\Lambda_\mathrm{th}}
\newcommand{\Lmax}{\Lambda_\mathrm{max}}
\newcommand{\xmax}{x_\mathrm{max}}
\newcommand{\xini}{\mathsf{x}_0}
\newcommand{\xfull}{x_\mathrm{full}}

\begin{document}

\title{Transverse Microwave Convective Instability at Transition Crossing}
\author{A.~Burov}
\email{burov@fnal.gov}
\affiliation{Fermilab, PO Box 500, Batavia, IL 60510-5011}
\date{\today}

\begin{abstract}

An analytical model of transverse convective instability of a bunch at transition crossing is presented for the microwave case, when the wake is sufficiently short compared with the bunch length. The space charge is assumed to be strong, as it typically is for the low and medium energy synchrotrons. A simple formula for the intensity threshold is obtained and its prediction is compared with available CERN PS data; without a single fitting parameter, a good agreement is demonstrated. 
 
\end{abstract}


\maketitle

 



\section{\label{sec:Int}Introduction}

Transverse convective instabilities may develop in a bunched beam when both the space charge tune shift and the wake-related coherent tune shift exceed the synchrotron tune~\cite{PhysRevAccelBeams.22.034202}. For a while, such a situation may take place for an accelerated beam at transition crossing, when the synchrotron tune crosses zero. However, the synchrotron tune remains small only for a limited time in this case; if this time is too short, the instability would not develop. 

Rising of computing power certainly allows shedding some light on this as well as on many other problems. However, the required macroparticle simulations, being valuable, are not sufficient for our needs, not only because they are still time-consuming, but also because, by themselves, they do not provide us with understanding of the underlying physics. They cannot provide us with proper concepts and relevant parameters, which we have to know before doing the simulations. That is why any attempt to build a reasonable analytical model of a complicated physical process, supported and followed by conceptual development, is and will remain indispensable. 

Led by this conviction, the author devised a model of a transverse microwave instability at transition crossing. The model, suggested in this paper, is not a first attempt in this direction, see e.g. Refs.~\cite{Gareyte:217745, Gareyte:477074, Cappi:2000ze}, but apparently it is a first one when the space charge is taken into account and shown to be important. It is also a first one demonstrating a good agreement with available observations. 

The model assumes rather simple bunch representation, just as a piece of a coasting beam at zero chromaticity, without gamma-jump. A resonator wake, short compared to the bunch length, is supposed. This choice of wake is motivated by the author's interest to compare his theory with observations at CERN PS accelerator, where such wake function is claimed to be close to the reality~\cite{PhysRevAccelBeams.21.120101}. Apart of its scientific importance, the PS transverse instability is of a special interest to the author who was issued a friendly challenge to apply his theory of convective instabilities to the PS~\cite{PhysRevAccelBeams.22.068002}. At the end of this paper, the theoretical result is compared with the measurements of Refs.~\cite{Kornilov:2013PS, PhysRevAccelBeams.21.120101}. Without a single fitting parameter, a good agreement is shown.

\section{\label{Sec:SSC} Strong Space Charge Approximation}

Let us assume that the SC transverse frequency shift sufficiently exceeds the band of coherent frequency shifts, which condition is usually well satisfied for low and medium energy proton accelerators. If so, the strong space charge approximation of Ref.~\cite{burov2009head, burov2015damping} may be applicable. Let us also be limited by a fast microwave instability, when the initial perturbation at the bunch center grows so fast that it does not reach the bunch tail for the relevant time. In this case, we may consider the bunch as a piece of a coasting beam which line density $\lambda=N/l_b$ and relative rms momentum spread $\delta p/p$ are the same as at the bunch center. For a real Gaussian-like bunch with maximal line density $\lambda_\mathrm{max}$, and rms length $\sigma_s$, this implies that $\lambda \simeq \lambda_\mathrm{max}$, while the effective bunch length has to be calculated as $l_b = \sqrt{2\pi} \sigma_s$. With these assumptions and definitions, the strong SC equation of motion can be written as follows,
\begin{equation}
i \frac{\pd x(s,t)}{\pd t} + \frac{v^2}{\omsc} \frac{\pd^2 x(s,t)}{\pd s^2}= -\frac{\lambda r_0 c}{4\pi \gamma \Qb} \int_0^s \dd s' W(s-s')x(s',t)\,.
\label{SSCs}
\end{equation}
Here $x(s,t)$ is an offset of a bunch slice at the intrabunch position $0\leq s \leq l_b$ at time $t$; $v$ is the rms spread of the effective longitudinal velocities $v=|\eta| c \delta p/p$, with $\eta$ as the slippage factor, $c$ as the speed of light; $\omsc$ is the transversely averaged SC frequency shift; $r_0$ is the proton classical radius; $\gamma$ is the Lorentz factor, and $\Qb$ is the betatron tune. The wake sign conventions assume its positive arguments and positive sign at small arguments.   

Generally speaking, Eq.~(\ref{SSCs}) requires initial and boundary conditions. However, if the bunch is long enough compared to the wake length, and the instability is sufficiently fast, the boundary conditions are irrelevant. In this paper we assume that such model requirements for the instability to be {\it microwave} are satisfied, checking at the end how well they are. We also neglect variations of the bunch length and momentum spread during the instability development, which is justified if the entire time of the process does not exceed few adiabatic times~\cite{lee2018accelerator}.

We suppose also that there are no external microwave excitations, and thus the initial conditions of the perturbation~(\ref{SSCs}) are provided solely by the Schottky noise. A problem remains though, which time to take as the initial; this problem is addressed in the following way. First, we solve Eq.~(\ref{SSCs}) assuming some initial conditions at the moment of the transition crossing, at $t=0$. The solution obtained in this way, as it will be shown below, has a form of a product of the small initial offset $\mathsf{x}_0$ and a presumably large exponent, i.e.$x(s,t)=\mathsf{x}_0 \exp(\Lambda(s,t))$. Thus, the total maximum, over $s$ and $t$, of the offset is expressed through the corresponding maximum of the exponent, $\xmax = \mathsf{x}_0 \exp(\Lmax)$. Such form of the result just shows that for the time $t >0$, the perturbation at $t=0$ is amplified by a factor of $\exp(\Lmax)$. It seems rather obvious to claim that the same amplification should happen for the time $t<0$. Thus, the full amplification of the initial conditions taken far before the transition crossing is $\exp(2\Lmax)$, and the maximal offset for the initial conditions $\mathsf{x}_0$ taken at, so to speak, $t=-\infty$, can be presented as $\xfull=\mathsf{x}_0 \exp(2\Lmax)$, where, let us repeat, $\Lmax$ is the maximal exponent for the case of the initial conditions taken at $t=0$. So, the entire recipe is following: first, we are solving the problem with the Schottky-noise initial conditions at $t=0$, and then just double the exponent.    

Now, after the complete description of the equation of motion, the boundary and initial conditions, we may proceed with the solution.    

After the Fourier transformation of Eq.~(\ref{SSCs}), with
\begin{equation}
x_k \equiv \int_0^{\infty}\dd s\, x(s) e^{-i k s}\,,
\label{Fourier}
\end{equation}
the equation reduces to the following form:
\begin{equation}
i \dot{x}_k - \frac{k^2 v^2}{\omsc} x_k = \Omk x_k\,, 
\label{SSCk}
\end{equation}
where the {\it dot} sign stands for the time derivative, and the coherent tune shift
\begin{equation}
 \Omk = -i \frac{\lambda r_0}{\gamma \Qb} \frac{Z_k}{Z_0}\,. 
\label{Omk}
\end{equation}
Here $Z_k$ is the transverse impedance, 
\begin{equation}
Z_k \equiv -i \int_0^{\infty}\dd s\, W(s) e^{-i k s}; \; \Im k < 0\,,
\label{Zk}
\end{equation}
and $Z_0=4\pi/c=377$~Ohm. 

In the case of transition crossing, near the transition energy, the effective velocity can be represented as
\begin{equation}
v=\frac{2 \dot{\gamma}}{\gamma^3} \frac{\delta p}{p} c\,t \equiv b\,t\,,
\label{velocity}
\end{equation}
assuming the transition is crossed at $t=0$; the parameter $b$ is an effective acceleration.  

Thus, with initial conditions $\left. x_k \right|_{t=0}=x_k(0)$, the solution of Eq.~(\ref{SSCk}) follows,
\begin{equation}
x_k(t)=x_k(0) e^{-i \Omk t -i B_k t^3/3};\; B_k \equiv k^2 b^2/\omsc \,.
\label{Solk}
\end{equation}
This, in turn, leads us, in the space domain, to 
\begin{equation}
x(s,t)=\int \frac{\dd k}{2\pi} x_k(0) e^{-i \Omk t +i k s - i B_k t^3/3}\,.
\label{Sols}
\end{equation}
In principle, this single integral solves the problem; the remaining part of the paper is devoted to its estimation. 

For the microwave instability, the typical initial offset is orders of magnitude smaller than the aperture, so our interest is essentially limited by a situation when the amplification of the initial perturbation is very large, i.e. the exponent of the integrand of Eq.~(\ref{Sols}) can be treated as big.    

Such sort of integrals can be effectively taken by the saddle-point method, see e.g. Ref.~\cite{bleistein1975asymptotic}. According to its recipe,   
\begin{equation}
\int_{-\infty}^{\infty}\frac{\dd k}{2\pi} f(k)\exp(-i \Phi(k)) \simeq f(k_0)\frac{\exp(-i \Phi(k_0))}{\sqrt{2\pi |\Phi''(k_0)|}}\,,
\label{SPM0}
\end{equation}
with $k_0$ as the saddle point in the complex plain of the wavenumber $k$, where the exponent derivative is zero, 
\begin{equation}
\Phi' \equiv \frac{\pd \Phi}{\pd k}=0
\label{SPEq}
\end{equation}
This asymptotical method is applicable if $|\Phi_{k_0}| \gg 1$ and the function $f(k)$ is smooth compared with the exponent.

\section{Short Resonant Wake}

Let us estimate the amplification of the initial perturbation~(\ref{Sols}) for a resonator wake 
\begin{equation}
W(s)=W_0 \sin(\kappa s)\exp(-\alpha s), 
\label{ResWakeDef}
\end{equation}
assuming $\alpha l_b \gg 1$.

For this wake function, the coherent tune shift $\Omk$~(\ref{Omk}) can be written as
\begin{equation}
\Omk=-\frac{w \kappa}{\kappa^2 -(k-i\alpha)^2}
\label{Omk2}
\end{equation}
with
\begin{equation}
w=\frac{\lambda r_0 c}{\gamma \Qb} \frac{R_s}{Q_r Z_0} \frac{\kappa^2+\alpha^2}{\kappa}\,,
\label{wpar}
\end{equation}
where $R_s$ is the so called shunt impedance and $Q_r=\sqrt{\kappa^2 + \alpha^2}/(2\alpha)$ is the quality factor.

The saddle-point integration with the phase of Eq.~(\ref{Sols})
\begin{equation}
\Phi(k)=\Omk t - k s + \frac{k^2 b^2 t^3}{3\, \omsc}\,
\end{equation}
is done in the Appendix, 
\begin{equation}
x(s,t) \simeq x_{\kappa}(0) \frac{(2w s t)^{1/4}}{2\sqrt{2\pi} s}\, \exp(-i \Psi(s,t))
\label{SolsFull}
\end{equation}  
where $\Psi(s,t) \equiv \Phi(k_0)$ is the saddle-point phase,
\begin{equation}
\Psi(s,t)= i\sqrt{2wt\left( s-\frac{2\kappa b^2 t^3}{3\omsc} \right)} -i\alpha s\ -\kappa s \,.
\label{SolsPsi}
\end{equation}  
For longitudinally frozen particles, $b=0$, it reduces to the Yokoya result~\cite{Yokoya:1986rk}. 

In the microwave case, when $\kappa l_b \gg 1$, and with absence of the external microwave noise, the initial conditions $x_{\kappa}(0)$ are determined by the Schottky noise in the bunch, so
\begin{equation}
x_{\kappa}(0) \simeq \frac{\sigma_x}{\sqrt{\lambda \kappa}} \,,
\label{Schottky}
\end{equation}
with $\sigma_x$ as the transverse rms size of the beam. 

Thus, with the logarithmic accuracy, the wave~(\ref{SolsFull}) can be presented as
\begin{equation}
x(s,t) \simeq \mathsf{x}_0 \exp(-i \Psi(s,t)),\;
\label{SolsFull2}
\end{equation}  
where
\begin{equation}
\mathrm{x}_0=\sigma_x/\sqrt{N\kappa l_b}\,
\label{x0}
\end{equation}  
is the effective initial offset at $t=0$. The crest of this wave reaches its maximum at $t=t_\mathrm{max}$, being as high as $\xmax=\xini e^{\Lmax}$, see Eqs.~(\ref{tmax}, \ref{smax}, \ref{Lmax}) of the Appendix.  

Assuming, as we discussed in Sec.~\ref{Sec:SSC}, that amplification of the Schottky perturbation before the transition is the same as it is after, the {\sf full} amplitude of the wave can be presented as
\begin{equation}
x_\mathrm{full}=\xini \exp(2\Lmax)\,.
\label{xfull}
\end{equation}  

The beam would be at least partially lost if at some time $t$ the initially small perturbation would grow to such an extent that the aperture limit $a$ is reached, i.e. the threshold of losses is determined by $\xfull = a$, or   
\begin{equation}
2 \Lmax = \ln(a/\xini) = \ln(a \sqrt{N\kappa l_b}/\sigma_x) \equiv \Lth\,. 
\label{thresh0}
\end{equation}
For CERN PS~\cite{PhysRevAccelBeams.21.120101}, the threshold logarithm is really large, $\Lth \approx 20$ for the entire range of typical parameters.  Substitution of Eq.~(\ref{Lmax}) yields the threshold condition as
\begin{equation}
\frac{w}{3\alpha^2 b} \sqrt{\frac{w\, \omsc}{\kappa}} = \Lth \,.
\label{thresh1}
\end{equation}
Conditions of applicability of this formula combine the microwave condition of the crest being within the bunch,
\begin{equation}
s_\mathrm{max}/c \leq \sigma_\tau\,,
\label{MWCond}
\end{equation}
the condition to be fast, $t_\mathrm{max} \leq (2-3) t_\mathrm{ad}$, where $t_\mathrm{ad}$ is the adiabatic time, and the strong space charge conditions, requiring for the SC frequency shift $\omsc$ to exceed both the maximal coherent tune shift $|\Omega_\kappa|=w/(2\alpha)$ and the maximal kinematic tune shift $\kappa v = \kappa b\,t_\mathrm{max}$,
\begin{equation}
\omsc \gg w/(2\alpha), \,   \kappa b\,t_\mathrm{max} \,.
\label{SSCcond}
\end{equation}

\section{Comparison with CERN PS}

To compare the computed threshold~(\ref{thresh1}) with observations, the relevant bunch parameters near the transition have to be available for the threshold intensity. 

As far as the author knows, there are two publications reporting such detailed measurements of the PS instability, a preprint of V.~Kornilov et al.~\cite{Kornilov:2013PS}, and more recent article of M.~Migliorati et al.~\cite{PhysRevAccelBeams.21.120101}. Let us start from the first of them.   

All the values we need can be extracted for two sets of the parameters reported in the preprint, namely, for the cases of $200$~kV and $110$~kV of the RF voltage. However, for $200$~kV the bunch was either longitudinally unstable or significantly mismatched: its maximal line density dropped about twice in the vicinity of the transition, while at $110$~kV the bunch profile was fairly stable. Due to these circumstances, we may use only the latter case for our purposes. For this case, the bunch of $N=(50 \pm 8)\cdot 10^{10}$ protons was reported to be at the loss threshold; its rms length $\sigma_\tau=\sigma_s/c$ can be estimated as $10$~ns, the relevant line density can be taken as $\lambda=N/(\sqrt{2\pi} \sigma_s)=6.6\cdot 10^8 \mathrm{cm}^{-1}$, and the relative rms momentum spread $\delta p/p=5.0\cdot 10^{-3}$, corresponding to the longitudinal rms emittance $0.3$eVs. The transverse normalized rms emittance was measured as $1.0\mathrm{\mu m}$, corresponding to the transversely averaged SC tune shift $0.075$. 

The PS vertical impedance is conventionally modeled by a broadband resonator with the shunt impedance $R_s=1.6\mathrm{M\Omega/m}$, centered at $0.8$GHz, with the Q-factor $Q_r=1.1$, according to Ref.~\cite{PhysRevAccelBeams.21.120101}.

Taking into account the PS energy ramp with $\dot{\gamma}=47\mathrm{s}^{-1}$ at the transition $\gamma=6.1$, the vertical tune $\Qb=6.3$, the mentioned above threshold exponent $\Lth=20$, the threshold bunch intensity determined by Eq.~(\ref{thresh1}) is calculated as $N=56\cdot 10^{10}$, within the error bars of the reported value of $(50 \pm 8)\cdot 10^{10}$. The growth time is computed as $t_\mathrm{max}=3.3$~ms, or about 2 PS adiabatic times; the crest travel distance $s_\mathrm{max}/c= 2w t_\mathrm{max}/(3\alpha^2 c)$ is found to be $8$~ns, a bit below the bunch rms length, being compatible with the microwave model. The strong space charge conditions~(\ref{SSCcond}) are also satisfied, $\omsc/|\Omega_\kappa| = 50$; $\omsc/(\kappa b\,t_\mathrm{max}) =7.$ 

\begin{figure}
\includegraphics[width=\linewidth]{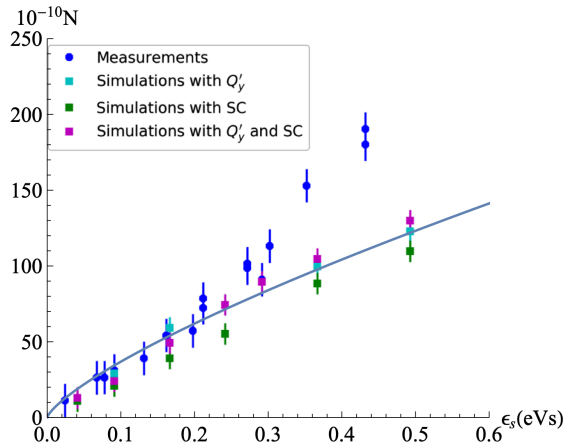}
\caption{\label{FigPS2}
Threshold number of protons versus longitudinal rms emittance for the PS bunch with the transverse rms emittance $5\mathrm{\mu m}$~\cite{PhysRevAccelBeams.21.120101}. The line corresponds to our formula~(\ref{thresh1}), the dots copy the data of Fig.~14 of Ref.~\cite{PhysRevAccelBeams.21.120101}  
	}
\end{figure}

To compare the model with the data of the Ref.~\cite{PhysRevAccelBeams.21.120101}, note that the threshold number of protons, $N=N_\mathrm{th}$, determined by Eq.~(\ref{thresh1}), scales with the longitudinal and transverse emittances, $\epsilon_s$ and $\epsilon_\perp$ correspondingly, as 
\begin{equation}
N_\mathrm{th} \propto \epsilon_s^{3/4} \epsilon_\perp^{1/4}\,.
\label{Nth}
\end{equation}  
Thus, for the reported transverse emittance $5\mathrm{\mu m}$, what is 5 times larger than in Ref.~\cite{Kornilov:2013PS}, the threshold intensity should be $5^{1/4}\cdot 56\cdot 10^{10}= 84\cdot 10^{10}$ protons per bunch, which lies within the measurement error bars. For this transverse emittance, the threshold bunch population of Eq.~(\ref{thresh1}) versus the longitudinal emittance is shown in Fig.~\ref{FigPS2}. The good agreement with the measurements worsens at larger emittances, $\epsilon_s \geq 0.3 \mathrm{\mu m}$, being still good with the tracking simulations. The author of this paper may suggest to check if the transverse emittance remains the same for larger longitudinal emittances. Since the SC tune shift grows along the line~(\ref{Nth}), one may suspect that the transverse emittance starts to grow as well, as it has been observed for larger SC tune shift of Ref.~\cite{Kornilov:2013PS}. 

A systematic difference between our model and tracking simulations at the level of $10-20\%$ in the whole range of the data of Fig.~\ref{FigPS2} would be even smaller, if to take into account that the threshold logarithm $\Lth$ for the simulations is about $30\%$ smaller than its real value, since the number of macroparticles is six orders of magnitude smaller than the real number of protons.

%

%

%
%

\section{Summary}

A transverse convective instability is sensitive both to the wake and space charge parameters, which are the related tune shifts versus the synchrotron tune~\cite{PhysRevAccelBeams.22.034202}. The higher the parameters, the stronger the instability. Thus, when the accelerated beam crosses the transition energy, it is maximally prone to this instability, which danger depends on how fast the transition is crossed. On the basis of strong space charge approximation~\cite{burov2009head, burov2015damping}, an analytical model of this instability is developed for a microwave case, when the wake is sufficiently short in comparison with the bunch length. The threshold condition is expressed by means of a simple formula~(\ref{thresh1}) without a single fitting parameter; a comparison with PS measurements~\cite{Kornilov:2013PS, PhysRevAccelBeams.21.120101} is undertaken and a good agreement is found.     

\begin{acknowledgments}

I am thankful to Mauro Migliorati for his help with understanding the measurements data reported in his article, and to Vladimir Kornilov for reminding me about his measurements and for discussing them. 

Fermilab is operated by Fermi Research Alliance, LLC under Contract No. DE-AC02-07CH11359 with the United States Department of Energy.

\end{acknowledgments}

\appendix*
\section{\label{sec:Saddle} Saddle Point Equation}

The saddle-point equation~(\ref{SPEq}), $\partial \Phi/\partial k=0$, has to be solved for
\begin{equation}
\Phi(k)=\Omk t - k s + \frac{k^2 b^2 t^3}{3\, \omsc}\,.
\label{Phi}
\end{equation}
With $k=\kbar \kappa +i\alpha$, the coherent frequency $\Omk$~(\ref{Omk2}) can be presented near the resonance $k\simeq \kappa$ as 
\begin{equation}
\Omk \approx \frac{w}{2\kappa} \frac{1}{\kbar -1}; \;\; \frac{d\Omk}{dk}=-\frac{wt}{2\kappa^2} \frac{1}{(\kbar-1)^2}\,.
\label{Omk3}
\end{equation}
Then, the saddle-point equation reduces to
\begin{equation}
\frac{1}{(\kbar-1)^2}=-\frac{2\kappa^2}{wt}\left(s-\frac{2\kappa b^2 t^3}{3\omsc}    \right)
\label{SPEq3}
\end{equation}
where we substituted $k \rightarrow \kappa$ in the acceleration-related term,  one more time assuming closeness to the resonance. Thus, at the saddle point, 
\begin{equation}
\Omk t = \frac{i}{2} \sqrt{2wt \left( s-\frac{2\kappa b^2 t^3}{3\omsc} \right)}
\label{Omkt}
\end{equation}
and the real part of the exponent
\begin{equation}
\Lambda (s,t) \equiv \Im\Phi_k=\sqrt{2wt\left( s-\frac{2\kappa b^2 t^3}{3\omsc} \right)} -\alpha s\,.
\label{ImPhikSP}
\end{equation}
Without acceleration, at $b=0$, this formula reduces to the Yokoya result~\cite{Yokoya:1986rk}. Hence, the wave crest $s_\star(t)$, where $\partial \Lambda/\partial s =0$, moves along the bunch according to 
\begin{equation}
s_\star(t) = \frac{wt}{2\alpha^2} + \frac{2\kappa b^2 t^3}{3\,\omsc}\,.
\label{crestsA}
\end{equation}
The crest exponent
\begin{equation}
\Lambda(s_\star,t) \equiv \Lambda_\star(t) = \frac{wt}{2 \alpha} - \frac{2\alpha \kappa b^2 t^3}{3\,\omsc}\,.
\label{LstarA}
\end{equation}
The exponent $\Lambda_\star(t)$ reaches its maximum at 
\begin{equation}
t=t_\mathrm{max}=\frac{1}{2\alpha b} \sqrt{\frac{w\, \omsc}{\kappa}}\,
\label{tmax}
\end{equation}
when the crest located at
\begin{equation}
s_\star(t_\mathrm{max}) \equiv s_\mathrm{max} = \frac{2w\,t_\mathrm{max}}{3\alpha^2}  =\frac{w}{3\alpha^3 b} \sqrt{\frac{w\, \omsc}{\kappa}}\,;
\label{smax}
\end{equation}
the exponent maximum is
\begin{equation}
\Lambda_\star(t_\mathrm{max}) \equiv \Lmax= \frac{w\,t_\mathrm{max}}{3\alpha}  =\frac{w}{6\alpha^2 b} \sqrt{\frac{w\, \omsc}{\kappa}}\,.
\label{Lmax}
\end{equation}

\bibliography{bibfile}			
\end{document}